\documentclass[
aps,prd,
12pt,
nopreprintnumbers,
showpacs,
eqsecnum,
nofootinbib
]{revtex4-1}

\usepackage{graphicx}
\usepackage{amssymb}

\begin{document}

\title{Geometrical effective actions for a partially massless spin-2 field}
\author{Nahomi Kan}\email[]{kan@gifu-nct.ac.jp}
\affiliation{National Institute of Technology, Gifu College,
Motosu-shi, Gifu 501-0495, Japan}
\author{Takuma Aoyama}\email[]{b014vbv@yamaguchi-u.ac.jp}
\affiliation{
Graduate School of Sciences and Technology for Innovation, Yamaguchi
University, Yamaguchi-shi, Yamaguchi 753--8512, Japan}
\author{Kiyoshi Shiraishi}\email[]{shiraish@yamaguchi-u.ac.jp}
\affiliation{
Graduate School of Sciences and Technology for Innovation, Yamaguchi
University, Yamaguchi-shi, Yamaguchi 753--8512, Japan}
\date{\today}
\begin{abstract}
We consider nonlinear effective actions for a spin-2 field, whose
`decoupling' limit gives Fierz--Pauli action in $D$ dimensional maximally
symmetric spacetime. We find, especially,  the effective action for a partially
massless field can take a concise geometrical form. The exact solution for time
evolution of the background metric in the model using the effective action is also
studied. 
\end{abstract}



\maketitle

\section{Introduction}
\label{introduction}
In the field-theoretical treatment, the gravitational interaction is mediated by a
massless graviton with spin 2.
There have been a lot of candidates for modifications of Einstein
gravity, motivated by the problems in modern cosmology, and the possibility of
massive gravitons has been proposed.
In their seminal paper \cite{FP}, Fierz and Pauli
presented an action for a free
 spin-2 field. It has been found that their linear equation of motion for a spin-2
field is unique in the flat spacetime, but there are difficulties both with the
formulation in the general curved spacetime \cite{AD1,AD2,BKP,BGKP,BGP} and with
the nonlinear extensions
\cite{BD1}.
Recently, consistent massive gravity and bigravity theories have been found and 
investigated \cite{Hinterbichler,deRham,SMvS}. The mass terms of symmetric tensor
fields in these theories have the complicated expressions of nonpolynomial and look
highly nonlinear. We should notice that the terms are not uniquely fixed and some
degree of arbitrariness still remains.

On the other hand, almost three decades ago, Grigoryan and Gottl\"ober presented a
nonlinear model of a symmetric second-rank tensor field in a beautiful, almost
unique, geometric form
\cite{GG1,GG2}; their main interest was in the introduction of a new field to
cause inflation of the Universe and they did not seem to consider mass of the
field seriously.

In the present paper, we extend their model to explore the classical action of a
massive spin-2 field in a maximally symmetric background spacetime.
We have found the action in a geometrically compact form, which coincides with the
action for a partially massless spin-2 field in a weak-field limit or
`decoupling' limit. 
Here the term `decoupling' means that nonlinear effects and possibly unphysical
modes are decoupled from the free theory in such a limit.%
\footnote{The use of the term also implies that the background
field is strongly curved.} 
Many difficulties are reported
in the nonlinear extension of the linear spin-2 theory, including some no-go
theorems, especially for partially massless fields
\cite{dHRT,GR}.  Although the model proposed in the present paper may not be
completely healthy, it would be worth studying as an effective field theory of a
complete theory including quantum gravity effects that will be revealed in the
future because there is a simple and interesting symmetry in the
model.

This paper is organized as follows.
In the next section, we review the spin-2 action in the 
spacetime manifold of constant curvature and its possible symmetries which emerge
at specific values of the mass parameter. The Grigoryan--Gottl\"ober (GG) theory,
which approximates the kinetic part of the spin-2 field theory, is reviewed in
Sec.~\ref{sec3}. In Sec.~\ref{sec4}, we examine the models for massive, massless,
and partially massless spin-2 fields in their decoupling limit, based on the
GG theory. It is shown that the action can be written in a
compact form when its decoupling limit describes the partially massless spin-2
field.  We investigate the exact solution for time evolution of the background
metric in the effective nonlinear theory with the partially massless spin-2 field
in Sec.~\ref{sec5}.  Finally, we summarize the results briefly and conclude in
Sec.~\ref{conclusion} with outlook.

\section{The Fierz--Pauli theory in the curved spacetime: a review}
\label{sec2}
In this section, we give an overview of the action for a free spin-2 field in a
curved background. The Fierz--Pauli action of a massive symmetric tensor field
$h_{\mu\nu}$ on a $D$-dimensional  Einstein spacetime with the constant
scalar curvature $R\equiv g^{\mu\nu}R_{\mu\nu}$, where the Ricci tensor
$R_{\mu\nu}$ equals to $\frac{R}{D}g_{\mu\nu}$, is written by
\cite{AD1,AD2,BKP,BGKP,BGP}
\begin{eqnarray}
S_{FP}&=&\int d^Dx\sqrt{-g}\Bigl[-\frac{1}{2}\nabla_\mu h_{\nu\lambda}\nabla^\mu
h^{\nu\lambda}+
\nabla_\nu h_{\mu\lambda}\nabla^\mu h^{\nu\lambda}-
\nabla_\mu h\nabla_\lambda h^{\mu\lambda}+
\frac{1}{2}\nabla_\mu h\nabla^\mu h\nonumber \\
& &+\frac{R}{D}\Bigl(h^{\mu\nu}h_{\mu\nu}-\frac{1}{2}h^2\Bigr)
-\frac{1}{2}m^2(h^{\mu\nu}h_{\mu\nu}-h^2)\Bigr]\,,
\label{FP}
\end{eqnarray}
where $h\equiv h^\rho_\rho=g^{\mu\nu}h_{\mu\nu}$, $\nabla_\mu$ represents the
covariant derivative in terms of the metric $g_{\mu\nu}$, and
$m$ is the mass parameter. The metric of the background geometry $g_{\mu\nu}$ is
assumed to be nondynamical, from this section to Sec.~\ref{sec4}. 

It is known that there are two special cases when the action becomes invariant
against the specific transformations.

If $m=0$, the action and the equation of motion are invariant under the
transformation
\begin{equation}
\delta
h_{\mu\nu}=\nabla_\mu\xi_\nu+\nabla_\nu\xi_\mu\,,
\end{equation}
where $\xi_\mu(x)$ is a vector gauge parameter.
By this symmetry, the degree of freedom of the symmetric tensor field $h_{\mu\nu}$
is reduced to be $D(D-3)/2$; the field describes a massless spin-2 particle.

If $m^2=\frac{D-2}{D(D-1)}R$, when the Higuchi bound \cite{Higuchi1,Higuchi2} is
saturated, the action takes the form
\begin{eqnarray}
S_{PM}&=&\int d^Dx\sqrt{-g}\Bigl[-\frac{1}{2}\nabla_\mu h_{\nu\lambda}\nabla^\mu
h^{\nu\lambda}+
\nabla_\nu h_{\mu\lambda}\nabla^\mu h^{\nu\lambda}-
\nabla_\mu h\nabla_\lambda h^{\mu\lambda}+
\frac{1}{2}\nabla_\mu h\nabla^\mu h\nonumber \\
&
&\qquad\qquad\qquad+\frac{R}{2D(D-1)}\left(Dh^{\mu\nu}h_{\mu\nu}-h^2\right)\Bigr]\nonumber
\\
&=&\int d^Dx\sqrt{-g}\Bigl[-\frac{1}{2}\nabla_\mu h_{\nu\lambda}\nabla^\mu
h^{\nu\lambda}+
\nabla_\nu h_{\mu\lambda}\nabla^\mu h^{\nu\lambda}-
\nabla_\mu h\nabla_\lambda h^{\mu\lambda}+
\frac{1}{2}\nabla_\mu h\nabla^\mu h\nonumber \\
&
&\qquad\qquad\qquad+\frac{m^2}{2(D-2)}\left(Dh^{\mu\nu}h_{\mu\nu}-h^2\right)\Bigr]\,.
\label{pm2}
\end{eqnarray}
Now, the action $S_{PM}$ is found to be invariant under the
deviation
\begin{equation}
\delta
h_{\mu\nu}=\Bigl(\nabla_\mu\nabla_\nu+\frac{m^2}{D-2}g_{\mu\nu}\Bigr)\zeta\,,
\end{equation}
where $\zeta(x)$ is a scalar gauge parameter.
The degree of freedom is then $D(D-1)/2-2$.
In this case, the field is called a partially massless field
\cite{DN1,DN2,DW2,DW3}.

\section{The GG action for a spin-2 field: a review}
\label{sec3}
In this section, we briefly review the GG theory of a
symmetric tensor field \cite{GG1,GG2}.
A symmetric tensor $f_{\mu\nu}=f_{\nu\mu}$ in the $D$-dimensional spacetime,
whose metric $g_{\mu\nu}$, is presumed.

First, we define 
\begin{equation}
\tilde{\Gamma}^\lambda_{\mu\nu}\equiv \frac{1}{2}f^{\lambda\sigma}\left(
\nabla_{\mu}f_{\sigma\nu}+\nabla_{\nu}f_{\sigma\mu}-
\nabla_{\sigma}f_{\mu\nu}\right)\,,
\end{equation}
where \underline{$f^{\mu\nu}$ is the inverse of $f_{\mu\nu}$}, i.e., $f^{\mu\sigma}
f_{\sigma\nu}=\delta^\mu_\nu$, and $\nabla_\mu$ denotes the covariant derivative in
terms of the metric $g_{\mu\nu}$. 
Note that the indices are lowered and raised by using the
background metric $g_{\mu\nu}$ and its inverse
$g^{\mu\nu}$, except for $f^{\mu\nu}$ which is the inverse of $f_{\mu\nu}$, i.e.,
$f^{\mu\nu}\ne g^{\mu\rho}g^{\nu\sigma}f_{\rho\sigma}$.
Note also that \underline{$\tilde{\Gamma}^\lambda_{\mu\nu}$ is a tensor}.
Interestingly, $\tilde{\Gamma}^\lambda_{\mu\nu}$ can be expressed in the form
\begin{equation}
\tilde{\Gamma}^\lambda_{\mu\nu}=\check{\Gamma}^\lambda_{\mu\nu}-
{\Gamma}^\lambda_{\mu\nu}\,,
\end{equation}
where
\begin{equation}
\check{\Gamma}^\lambda_{\mu\nu}=\frac{1}{2}f^{\lambda\sigma}\left(
\partial_{\mu}f_{\sigma\nu}+\partial_{\nu}f_{\sigma\mu}-
\partial_{\sigma}f_{\mu\nu}\right)\,,
\end{equation}
and ${\Gamma}^\lambda_{\mu\nu}$ is the usual Christoffel symbol (constructed from
$g_{\mu\nu}$).

Next, using this tensor, we define the curvature-like tensor
\begin{equation}
\tilde{R}^\lambda{}_{\rho\mu\nu}\equiv\nabla_\mu\tilde{\Gamma}^\lambda_{\nu\rho}
-\nabla_\nu\tilde{\Gamma}^\lambda_{\mu\rho}+
\tilde{\Gamma}^\lambda_{\mu\sigma}\tilde{\Gamma}^\sigma_{\rho\nu}
-\tilde{\Gamma}^\lambda_{\nu\sigma}\tilde{\Gamma}^\sigma_{\rho\mu}\,.
\end{equation}
Similarly, one can find
\begin{equation}
\tilde{R}^\lambda{}_{\rho\mu\nu}=\check{R}^\lambda{}_{\rho\mu\nu}-
{R}^\lambda{}_{\rho\mu\nu}\,,
\end{equation}
where
\begin{equation}
\check{R}^\lambda{}_{\rho\mu\nu}\equiv\partial_\mu\check{\Gamma}^\lambda_{\nu\rho}
-\partial_\nu\check{\Gamma}^\lambda_{\mu\rho}+
\check{\Gamma}^\lambda_{\mu\sigma}\check{\Gamma}^\sigma_{\rho\nu}
-\check{\Gamma}^\lambda_{\nu\sigma}\check{\Gamma}^\sigma_{\rho\mu}\,,
\end{equation}
and ${R}^\lambda{}_{\rho\mu\nu}$ is the usual Riemann tensor (constructed from
the Christoffel symbol ${\Gamma}^\lambda_{\mu\nu}$).
We also consider the symmetric tensor from the contraction of the tensor defined
above:
\begin{equation}
\tilde{R}_{\mu\nu}\equiv\tilde{R}^\rho{}_{\mu\rho\nu}
=\check{R}_{\mu\nu}-{R}_{\mu\nu}\,,
\end{equation}
where $\check{R}_{\mu\nu}=\check{R}^\rho{}_{\mu\rho\nu}$, and
$R_{\mu\nu}={R}^\rho{}_{\mu\rho\nu}$ is the usual Ricci tensor (constructed from
the Christoffel symbol ${\Gamma}^\lambda_{\mu\nu}$).
We adopt the action with rather geometrical form at last:
\begin{equation}
S_{GG}=-\frac{2}{\varepsilon^2}\int d^Dx \sqrt{-g}g^{\mu\nu}\tilde{R}_{\mu\nu}\,,
\end{equation}
where $\varepsilon$ is a parameter.
When we represent the small
fluctuation of $f_{\mu\nu}$ from $g_{\mu\nu}$ by $\psi_{\mu\nu}$, 
using the small parameter $\varepsilon$, i.e.,
\begin{equation}
f_{\mu\nu}=g_{\mu\nu}+\varepsilon \psi_{\mu\nu}\,,
\end{equation}
we find
\begin{eqnarray}
& &\lim_{\varepsilon\rightarrow
0}S_{GG}=-\lim_{\varepsilon\rightarrow
0}\frac{2}{\varepsilon^2}\int
d^Dx\sqrt{-g}g^{\mu\nu}\left(\tilde{\Gamma}^\rho_{\rho\sigma}\tilde{\Gamma}^\sigma_{\mu\nu}
-\tilde{\Gamma}^\rho_{\mu\sigma}\tilde{\Gamma}^\sigma_{\rho\nu}\right)\nonumber \\
& &=\int d^Dx\sqrt{-g}\left[-\frac{1}{2}\nabla_\mu\psi_{\nu\lambda}
\nabla^\mu\psi^{\nu\lambda}+
\nabla_\nu\psi_{\mu\lambda}\nabla^\mu\psi^{\nu\lambda}-
\nabla_\mu\psi\nabla_\lambda\psi^{\mu\lambda}+
\frac{1}{2}\nabla_\mu\psi\nabla^\mu\psi\right]\,,
\end{eqnarray} 
where $\psi\equiv\psi^\mu_\mu=g^{\mu\nu}\psi_{\mu\nu}$ and the boundary terms are
omitted. We can easily verify this equation if we utilize the expressions
\begin{equation}
\tilde{\Gamma}^\lambda_{\mu\nu}=\frac{\varepsilon}{2}g^{\lambda\sigma}\left(
\nabla_{\mu}\psi_{\sigma\nu}+\nabla_{\nu}\psi_{\sigma\mu}-
\nabla_{\sigma}\psi_{\mu\nu}\right)+O(\varepsilon^2)\,,
\end{equation}
and
\begin{equation}
g^{\mu\nu}\tilde{R}_{\mu\nu}=g^{\mu\nu}
\left(\tilde{\Gamma}^\rho_{\rho\sigma}\tilde{\Gamma}^\sigma_{\mu\nu}
-\tilde{\Gamma}^\rho_{\mu\sigma}\tilde{\Gamma}^\sigma_{\rho\nu}\right)+
\nabla_\rho\left(g^{\mu\nu}\tilde{\Gamma}^\rho_{\mu\nu}-
g^{\rho\lambda}\tilde{\Gamma}^\sigma_{\lambda\sigma}\right)\,.
\end{equation}

Since the limit of $\varepsilon\rightarrow 0$ yields the linear theory, 
we call the limit as the
`decoupling' limit, in which the nonlinear contributions decouple from the free
theory.%
\footnote{Note that the nonlinearity in terms of $\psi_{\mu\nu}$ lurks in
$f^{\mu\nu}$.}
 This is, of course, equivalent to the weak-field limit in the classical
theory. 

The action $S_{GG}$  supplies only the kinetic part for the symmetric tensor
field in the decoupling limit.
It coincides with the form of the kinetic part of the Fierz--Pauli action, or that
in the flat spacetime.
 In the next section, we examine the remaining quadratic terms in the tensor field
without derivatives. 

Lastly, we should remember the invariance of the action $S_{GG}$ under the
\underline{global} scale transformation \cite{GG2}%
\footnote{Of course, the reflection symmetry under $f_{\mu\nu}\leftrightarrow
-f_{\mu\nu}$ also exists.}
\begin{equation}
f_{\mu\nu}\rightarrow \Omega_f^2 f_{\mu\nu}\,,
\label{gst}
\end{equation}
where $\Omega_f$ is a constant.

\section{construction of massive, massless, and partially massless spin-2 theories
based on the GG theory}
\label{sec4}

\subsection{general massive case}
Now, we search the extended version of the GG theory whose decoupling limit
coincides with the massive Fierz--Pauli theory in the curved spacetime.
We can construct arbitrary quadratic terms without derivatives of the field and add
them to the kinetic term in the GG action, which has been shown in the previous
section. For example, we can use the combinations of
\begin{equation}
\varepsilon^2\psi^2=(g^{\mu\nu}f_{\mu\nu}-D)^2\,,\quad
\varepsilon^2\psi_{\mu\nu}\psi^{\mu\nu}=g^{\mu\rho}g^{\nu\sigma}(f_{\mu\nu}-g_{\mu\nu})
(f_{\rho\sigma}-g_{\rho\sigma})\,.
\end{equation}
Notice that the quadratic terms in $\psi_{\mu\nu}$ can also arises from
\begin{equation}
g_{\mu\nu}f^{\mu\nu}=D-\varepsilon\psi+\varepsilon^2\psi^{\mu\nu}\psi_{\mu\nu}+
O(\varepsilon^3)\,.
\end{equation}

Because arbitrariness is too large, we will stop pursuing the fixing the
form of the generic quadratic term without derivatives until new findings are made
on its form.

\subsection{massless case}

Now, we are aware of the following relation:
\begin{equation}
\sqrt{|\det f_{\mu\nu}|}=\sqrt{|f|}=\sqrt{-g}\left[1+\frac{\varepsilon}{2}\psi
-\frac{\varepsilon^2}{4}\left(\psi^{\mu\nu}\psi_{\mu\nu}-\frac{1}{2}\psi^2\right)+
O(\varepsilon^3)\right]\,.
\end{equation}
Although its quadratic part is proportional to that with massless Fierz--Pauli
Lagrangian, the linear term $\propto \psi$ remains and the subtraction of it is
needed.
Rather, it may be possible to adopt the Born--Infeld-type gravitational
action \cite{BHOR}, inspired by the use of square roots. That is:
\begin{eqnarray}
& &S_{BI}\sim-\int d^Dx\Bigl[\sqrt{g_{\mu\nu}+\alpha(f_{\mu\nu}-g_{\mu\nu})+\beta
\tilde{R}_{\mu\nu}+\gamma g_{\mu\nu}g^{\rho\sigma}\tilde{R}_{\rho\sigma}}\nonumber
\\ & &\qquad\qquad\qquad\qquad+
\sqrt{g_{\mu\nu}-\alpha(f_{\mu\nu}-g_{\mu\nu})+\beta'
\tilde{R}_{\mu\nu}+\gamma'
g_{\mu\nu}g^{\rho\sigma}\tilde{R}_{\rho\sigma}}-2\sqrt{-g}\Bigr]\,,
\end{eqnarray}
where $\alpha$, $\beta$, $\gamma$, $\beta'$, and $\gamma'$ are constants.
Although this form seems lengthy, nonpolynomial, and less geometrical, this form
has its own unique interests. If we replace
$\tilde{R}_{\mu\nu}\rightarrow
\tilde{\Gamma}^\rho_{\rho\sigma}\tilde{\Gamma}^\sigma_{\mu\nu}
-\tilde{\Gamma}^\rho_{\mu\sigma}\tilde{\Gamma}^\sigma_{\rho\nu}$,%
\footnote{Please remember that $\tilde{\Gamma}^\lambda_{\mu\nu}$ is a tensor}
the dynamical model is apparently free from the Ostrogradsky instability
\cite{Ostrogradsky}, because there is no second derivatives in the time
coordinate. However, on the other hand, there is also a disadvantage that there is
room to select the second-order terms in $\varepsilon$ in the square root.
In addition, the original scale invariance (\ref{gst}) of $S_{GG}$ is broken in
this type of action, so we will leave the exploration around here as well.

\subsection{partially massless case}
The combination
\begin{equation}
(g_{\rho\sigma}f^{\rho\sigma})(g^{\mu\nu}f_{\mu\nu})=D^2+\varepsilon^2\left(D\psi^{\mu\nu}\psi_{\mu\nu}-
\psi^2\right)+
O(\varepsilon^3)\,,
\end{equation}
is scale invariant under the global transformation (\ref{gst}),
and the quadratic part in $\psi_{\mu\nu}$ is proportional to that of the Lagrangian
for a partially massless spin-2 field (\ref{pm2}).%
\footnote{Incidentally, the additional cubic and quartic terms in the theory in
$D=4$ suggested by de Rham
\textit{et al.} (found in Sec.~III C of Ref.~\cite{dHRT}) is invariant under the
scale transformation of their metric tensor.} Alternatively, in the Einstein
background metric, i.e.,
$R_{\mu\nu}=\frac{R}{D}g_{\mu\nu}$, we find
\begin{equation}
R_{\rho\sigma}f^{\rho\sigma}g^{\mu\nu}f_{\mu\nu}=\frac{R}{D}
(g_{\rho\sigma}f^{\rho\sigma})(g^{\mu\nu}f_{\mu\nu})=\frac{R}{D}
\left[D^2+\varepsilon^2\left(D\psi^{\mu\nu}\psi_{\mu\nu}-
\psi^2\right)+
O(\varepsilon^3)\right]\,.
\end{equation}
Further removing the restriction and assuming maximally symmetric spacetime,
such as the de Sitter spacetime,
where $R_{\mu\nu\rho\sigma}=\frac{R}{D(D-1)}(g_{\mu\rho}g_{\nu\sigma}-
g_{\mu\sigma}g_{\nu\rho})$,
we have
\begin{eqnarray}
& &f_{\mu\nu}R^\mu{}_\rho{}^\nu{}_\sigma f^{\rho\sigma}=\frac{R}{D(D-1)}
\left(g^{\mu\nu}f_{\mu\nu}
g_{\rho\sigma}f^{\rho\sigma}-D\right)\nonumber \\
& &=\frac{R}{D(D-1)}
\left[D(D-1)+\varepsilon^2\left(D\psi^{\mu\nu}\psi_{\mu\nu}-
\psi^2\right)+
O(\varepsilon^3)\right]\,.
\end{eqnarray}

From the above facts, we can propose the following candidates for the
pre-decoupled action of the part that does not include the derivative of the field
$f_{\mu\nu}$:

\begin{itemize}
\item Type-$I$
\begin{equation}
\Delta S_{PMI}=\frac{1}{2\varepsilon^2}\int
d^Dx\sqrt{-g}\,(f_{\mu\nu}R^\mu{}_\rho{}^\nu{}_\sigma f^{\rho\sigma}-R)
\end{equation}
\item Type-$II$
\begin{equation}
\Delta S_{PMII}=\frac{1}{2(D-1)\varepsilon^2}\int
d^Dx\sqrt{-g}\,(R_{\rho\sigma}f^{\rho\sigma}g^{\mu\nu}f_{\mu\nu}-DR)
\end{equation}
\item Type-$III$
\begin{equation}
\Delta S_{PMIII}=\frac{1}{2D(D-1)\varepsilon^2}\int
d^Dx\sqrt{-g}\,R\,(g_{\rho\sigma}f^{\rho\sigma}g^{\mu\nu}f_{\mu\nu}-D^2)
\end{equation}
\item Type-$III'$
\begin{equation}
\Delta S_{PMIII'}=\frac{\Lambda}{(D-1)(D-2)\varepsilon^2}\int
d^Dx\sqrt{-g}\,(g_{\rho\sigma}f^{\rho\sigma}g^{\mu\nu}f_{\mu\nu}-D^2)
\end{equation}
\end{itemize}

Note that Type-$III'$ is merely Type-$III$ with the replacement
$R=\frac{2D}{D-2}\Lambda$, where $\Lambda$ is a constant.
Note that Type-$I$, Type-$II$, and Type-$III$ is transformed as
\begin{equation}
\Delta S_{PM*}\rightarrow \Omega_g^{D-2} \Delta S_{PM*}\qquad
(*=I, II, III)
\end{equation}
under the global scale
transformation of the background geometry
\begin{equation}
g_{\mu\nu}\rightarrow \Omega_g^2 g_{\mu\nu}\,,
\end{equation}
but Type-$III'$ is transformed as $\Delta S_{PMIII'}\rightarrow \Omega_g^{D}
\Delta S_{PMIII'}$. Note also that the kinetic part $S_{GG}$ is transformed as
$S_{GG}\rightarrow
\Omega_g^{D-2} S_{GG}$.
Of course, a linear combination of these can be the mass term of the partially
massless field in the decoupling limit. 

If and only if we consider the fluctuation of the background metric
$\delta g_{\mu\nu}$, then the nonlinear couplings to the field
$\psi_{\mu\nu}$ beyond the decoupling limit appear as in bigravity models, and we
can distinguish the different types of the mass term.
To select a unique mass term by the difference at the fully nonlinear level,
we need to consider an extended partially massless condition.
However, it is difficult to analyze the extention considering general
nonlinear and derivative couplings as well as bigravity%
\footnote{Note also that the kinetic terms for $f_{\mu\nu}$ and $g_{\mu\nu}$ in the
present model are not symmetric unlike those in general bigravity models.},
and therefore it is one of the future tasks.

Note that only Type-$I$ is independent of the
spatial dimension (up to spacetime integration), and any other linear combinations
cannot be independent of the dimension of the spacetime.


Before closing this section, we place a comment on another possible combination
which leads to the quadratic term of partially massless spin-2 theory in the
decoupling limit. That is: 
\begin{equation}
\sqrt{|f|}(g_{\mu\nu}f^{\mu\nu})^{D/2}=D^{D/2}\sqrt{-g}\left[1
+\frac{\varepsilon^2}{4}\left(\psi^{\mu\nu}\psi_{\mu\nu}-\frac{1}{D}\psi^2\right)+
O(\varepsilon^3)\right]\,.
\end{equation}
Also from this fact, we can guess that the scale invariance in terms of
$f_{\mu\nu}$ is closely related to the partially masslessness in the decoupling
limit.


\section{A little bit cosmology: An exact solution for time evolution of the
background metric %
}
\label{sec5}
In this section, we shall consider cosmology based on the model in honor of the
first spirit of Grigoryan and Gottl\"ober.
This time, we examine the last model in the previous section, that is, the
effective nonlinear theory with a partially massless field as the decoupling limit.

We add the Einstein--Hilbert action with the cosmological constant $\Lambda$ so
that the background spacetime can be maximally symmetric.
Then, de Sitter spacetime is expected to be realized asymptotically.
Therefore, we consider the following action:
\begin{equation}
S=S_{GG}+\Delta S_{PMI}+\int
d^Dx\sqrt{-g}\Bigl[R-2\Lambda\Bigr]\,.
\label{feq}
\end{equation}

Needless to say, it is hard to find exact solutions of the equations derived from
the action (\ref{feq}). In bigravity theories, the strong ansatz that 
two metrics are proportional to each other, i.e.,
$f_{\mu\nu}=c^2g_{\mu\nu}$, is adopted to seek exact solutions in several papers
\cite{SMvS,HSS1,HSS2,KN,Katsuragawa}. Based on that, we here adopt the ansatz
\begin{equation}
f_{\mu\nu}=e^{\sqrt{2}\varepsilon\phi(x)}g_{\mu\nu}\,.
\end{equation}
This is the slimmest assumption; under this condition, $\Delta S_{PMI}$
vanishes (also, $\Delta S_{PMII}=\Delta S_{PMIII}=\Delta S_{PMIII'}=0$) because
of the scale invariance under (\ref{gst}).
On the other hand, the kinetic part $S_{GG}$ results in
\begin{equation}
S_{GG}\rightarrow+\int
d^Dx\sqrt{-g}\Bigl[(D-1)(D-2)g^{\mu\nu}\partial_\mu\phi
\partial_\nu\phi\Bigr]\,,
\end{equation}
which corresponds to the action for a
massless phantom scalar field \cite{Caldwell,CKW,DSS,DKS}, that is, has the
opposite sign to the action for the canonical scalar field.

Now, let us go for the exact solution.
We assume the following metric:
\begin{equation}
g_{\mu\nu}dx^\mu dx^\nu=-dt^2+a^2(t)\delta_{ij}dx^idx^j\,,
\end{equation}
where $a(t)$ is the scale factor, $t=x^0$, and $i=1,2,3$.
Accordingly, we also assume that $\phi$ is homogeneous and expressed as a function
of $t$.
Then, the Friedmann equation is expressed as
\begin{equation}
\left(\frac{\dot{a}}{a}\right)^2=H_0^2-\dot{\phi}^2\,,
\label{fe}
\end{equation}
where the dot $(\dot{~})$ denotes the time derivative and
$H_0^2\equiv\frac{2\Lambda}{(D-1)(D-2)}$.        
In addition, $\phi$ obeys the equation of motion
\begin{equation}
\ddot{\phi}+(D-1)\frac{\dot{a}}{a}\dot{\phi}=0\,.
\end{equation}
From the first integration of this equation, one obtain 
$\dot{\phi}^2=C^2/a^{2D-2}$ with a constant $C$.
Now, one finds, from (\ref{fe})
\begin{equation}
\dot{V}=(D-1)H_0\sqrt{V^2-V_0^2}\,,
\end{equation}
with $V\equiv a^{D-1}$, $V_0\equiv C/H_0$,
which can easily be integrated to yield
\begin{equation}
V(t)=a^{D-1}(t)=V_0\cosh[(D-1)H_0(t-t_0)]\,,
\end{equation}
where $t_0$ is an integration constant.
With this solution, we obtain
\begin{equation}
\phi(t)=\phi(t_0)+\frac{2}{D-1}\arctan\tanh\frac{(D-1)H_0(t-t_0)}{2}\,.
\end{equation}
Note that $\lim_{t\rightarrow\infty}\phi(t)=\phi(t_0)+\frac{\pi}{2(D-1)}$.

This analytic solution implies that the background geometry asymptotically
approaches de Sitter spacetime, i.e., the scale factor behaves like $a(t)\sim
e^{H_0 t}$ as
$t\rightarrow\infty$ and, at the same time, the scalar mode $\phi(t)$ of the tensor
field $f_{\mu\nu}$  rapidly approaches a constant. Recall that the systems
connected by the global scale transformation (\ref{gst}) are equivalent. As for an
excitation mode, $\phi$ is considered to be identified with a decoupled degree
of freedom,
 so this result can be interpreted as a mechanism by which the
nonphysical mode freezes dynamically.
 However, our true cosmological interest is undoubtedly the
contribution of the spin-2 field as a matter field to the development of the
Universe. 
For this purpose, we should try to include some tensor modes by adopt more
general ansatz such as
$f_{\mu\nu}=e^{\sqrt{2}\epsilon\phi}(g_{\mu\nu}+\psi\delta_{\mu 0}\delta_{\nu 0})$ 
etc., though no analytic solution can be found (using either mass term).
In this paper, we only showed analytically the most important character discussed
so far, and we would like to investigate generic features in cosmology with
numerical methods to obtain definite results in future separate papers.

\section{Summary and conclusion}
\label{conclusion}
In this paper, we have studied effective field theories of a spin-2 field,
especially one of which the decoupling limit yields the partially massless spin-2
field theory. It has a simple geometrical form and has the symmetry under
the global scale invariance.
For instance, we considered the action
\begin{equation}
S_{GGPMI}=S_{GG}+\Delta S_{PMI}=\frac{1}{\varepsilon^2}\int d^Dx
\sqrt{-g}\Bigl[-2g^{\mu\nu}\tilde{R}_{\mu\nu}
+\frac{1}{2}(f_{\mu\nu}R^\mu{}_\rho{}^\nu{}_\sigma f^{\rho\sigma}-R)\Bigr]\,,
\end{equation}
and if the background geometry admits maximal symmetry, one finds
\begin{eqnarray}
& &\lim_{\varepsilon\rightarrow
0}S_{GGPMI}\nonumber \\
& &=\int
d^Dx\sqrt{-g}\left[-\frac{1}{2}\nabla_\mu\psi_{\nu\lambda}
\nabla^\mu\psi^{\nu\lambda}+
\nabla_\nu\psi_{\mu\lambda}\nabla^\mu\psi^{\nu\lambda}-
\nabla_\mu\psi\nabla_\lambda\psi^{\mu\lambda}+
\frac{1}{2}\nabla_\mu\psi\nabla^\mu\psi\right]\nonumber \\
&
&\qquad\qquad\qquad+\frac{R}{2D(D-1)}\left(D\psi^{\mu\nu}\psi_{\mu\nu}
-\psi^2\right)\Bigr]\,,
\end{eqnarray}
where $\psi_{\mu\nu}=\varepsilon^{-1}(f_{\mu\nu}-g_{\mu\nu})$.

The nonlinear terms in the expansion of our action may include unphysical
modes for a finite $\varepsilon$. 
Our model is well worth studying for brave researchers who tolerate phantom
fields in cosmology.%
\footnote{It is interesting to extend our model to include bouncing cosmologies
\cite{NB} with the phantom degrees of freedom.} Unfortunately, this model as
bigravity does not fall into Hassan--Rosen class
\cite{SMvS,HR2} of the bimetric theory%
\footnote{Still, we would like to point out that GG theory is not a just a double
copy of general relativity.}
 and therefore will have instability in the
whole theory with finite $\varepsilon$. 
Elimination of unstable degrees of freedom will require further ingenuity in
$O(\varepsilon)$.
Further, it would be necessary to consider higher order terms
and derivative terms of the symmetric tensor with a view to symmetry.
Additional auxiliary degrees of freedom may also be required.
Therefore, if it is necessary to append an infinite number of additional terms in
the action, we should rather consider the deep relationship with string theory
\cite{LMMS}. 
It is also known that the bimetric theories play an important role in quantum
gravity with the functional renormalization group equation \cite{MRS}. 
Bimetric theories of gravity has been studied also by models in the first order
formalism \cite{AS} or in the Plebanski theory \cite{Speziale}, which is
closely related to spin form models \cite{Perez1,Perez2} and thus to the loop
quantum cosmology \cite{AshSin}. In any case, we hope that our model will provide
a clue to the quantum spin-2 or quantum gravity theory.

We presented several candidates for terms with no derivative on the field,
which lead to the same decoupling limit.
Among these, $\Delta S_{PMI}$ has no explicit description of the dimension $D$.
Therefore, it is expected that partial masslessness would be inherited with
$S_{GGPMI}$ even if the dimensional reduction or the Kaluza--Klein
compactification of the background metric is performed.
The first candidate for spacetime to consider would be the Nariai spacetime
\cite{Nariai} and the generalized higher-dimensional Nariai spacetimes.
In contrast, the study of lower dimensional models is also interesting.

Finally, we list some other possible future explorations as follows. 
Further generalizations can be obtained by considering
the mixing terms with the massless graviton and couplings between the tensor and
other matter fields. It would be interesting to consider supersymmetric
and/or Born--Infeld extension of the present model. Finding solutions for
compact objects including black holes and wormholes
 is also an interesting topic. Furthermore, a study of the theory in general
background and the quantum cosmological treatment of the theory would be equally
worth pursuing.

\bibliographystyle{apsrev4-1}


\end{document}